\pdfoutput=1
\documentclass[prb,twocolumn,a4paper,superscriptaddress,floatfix,showpacs]{revtex4}
\usepackage{graphicx}
\usepackage{amssymb}
\usepackage{amsmath}
\usepackage{color}
\usepackage{paralist}
\usepackage{bm}
\usepackage{verbatim}

\def\cH{\hat{\cal H}}
\def\cL{{\cal L}}
\def\cR{{\cal R}}

\def\cW{{\cal W}}

\def\bA{{\bf A}}

\def\bF{{\bf F}}

\def\br{{\bf r}}

\def\bp{{\bf p}}
\def\bv{{\bf v}}

\def\hbsigma{\hat{\boldsymbol \sigma}}

\def\hV{\hat V}

\def\hsigma{\hat\sigma}

\newcommand{\ket}[1]{| #1 \rangle}

\def\vf{\varphi}
\def\de{\partial}

\def\br{\mathbf{r}}

\def\bp{\mathbf{p}}

\def\bv{\mathbf{v}}

 1

\begin{document}



\title{Strongly anisotropic Dirac quasiparticles in irradiated graphene}


\author{S.V.~Syzranov} \affiliation{Institute for Theoretical
Condensed Matter Physics, Karlsruhe Institute of Technology, 76131
Karlsruhe, Germany}

\author{Ya.I.~Rodionov} \affiliation{Institute for
Theoretical and Applied Electrodynamics RAS,
125412 Moscow, Russia}

\author{K.I.~Kugel} \affiliation{Institute for
Theoretical and Applied Electrodynamics RAS,
125412 Moscow, Russia} \affiliation{Center of Emergent
Matter Science, RIKEN, Saitama, 351-0198, Japan}

\author{F.~Nori} \affiliation{Center of Emergent Matter
Science, RIKEN, Saitama, 351-0198, Japan} \affiliation{Department
of Physics, University of Michigan, Ann Arbor, MI 48109-1040, USA}

\date{\today}

\begin{abstract}
We study quasiparticle dynamics in graphene exposed to a
linearly-polarized electromagnetic wave of very large intensity.
Low-energy transport in such system can be described by an effective time-independent Hamiltonian, characterized by multiple Dirac points in the first Brillouin zone. Around each Dirac point the spectrum is anisotropic: the velocity along the polarization of the radiation significantly exceeds the velocity in the perpendicular direction. Moreover, in some of the points the transverse velocity oscillates as a function of the radiation intensity. We find that the conductance of a graphene p-n junction in the regime of strong irradiation depends on the polarization as $G(\theta)\propto|\sin\theta|^{3/2}$, where $\theta$ is the angle between the polarization and the p-n interface, and oscillates as a function of the
radiation intensity.
\end{abstract}

\pacs{78.67.Wj, 05.60.Gg, 72.20.Ht, 72.80.Vp}



\maketitle

{\it Introduction.} Exposing conducting materials to strong monochromatic radiation may reveal many fundamental effects in their optical and transport properties. For instance, external
radiation has been predicted to open dynamical gaps in
the quasiparticle spectra in clean semiconductors~\cite{Galitskii:firstgap},
turning conducting materials into insulators~\cite{GoreslavskiiElesin:metalinsulator},
strongly changing the coefficient of light absorption~\cite{AlexandrovElesin:absorption}, and leading to interference transport phenomena in the presence of a coordinate-dependent potential~\cite{Fistul:interference}.
Another fascinating example is the radiation-induced Hall effect~\cite{OkaAokiHall:Hall} occurring in the case of a circularly polarized electromagnetic wave.

Although it is now more than 40 years since radiation-induced dynamical gaps were predicted~\cite{Galitskii:firstgap},
their manifestations have never been detected in transport phenomena, yet spectroscopic observations of the dynamical gaps are very few~\cite{Elesin:firstgapobservation,Vu:Mollowtriplet}.
Since interactions and disorder in realistic materials obscure coherent radiation-induced effects, it requires enormous radiation intensities to observe modifications of the quasiparticle spectra by radiation.

The strength of the resonant interaction between the charge carriers and electromagnetic field 
can be characterized by the parameter
$\alpha(\omega)=|e|Ev/\omega^2\equiv\Delta/\omega$,
where $E$ is the amplitude of the field, $v$ is the characteristic charge carrier velocity, $\omega$ is the frequency of the field, and $\Delta$ is the value of the dynamical gap in the absence of disorder and interactions, $\hbar=1$. For the resonant interaction in a conventional semiconductor, the maximum value of $\alpha$ for a given radiation intensity is determined by the minimal possible frequency, i.e. by the width of the band gap. In fact, until recently, attainable light intensities would allow one to access only the weak-interaction regime, $\alpha(\omega)\ll1$, corresponding to a weak modification of the charge carrier spectra, except maybe for a very small vicinity of the dynamical gaps.

The situation might have changed recently with the advent of new semiconductors without band gaps: graphene and the surfaces of topological insulators in higher dimensions can be treated as semiconductors with touching conduction and valence bands, quasiparticle dynamics being described by effective Dirac equations near the respective points (Dirac points).
The absence of the gap between the conduction and the valence bands allows one to access the regime of strong interaction [$\alpha(\omega)\gg1$] between the charge carriers and external electromagnetic field by applying radiation of very low frequency $\omega$ rather than by using large radiation intensities.

How would such interaction affect the charge carrier dynamics?
Does it have any observable manifestations for realistic strengths of disorder and electron-electron interactions?

In this paper, we study quasiparticle dynamics in graphene subject to a strong linearly polarized electromagnetic field. We demonstrate that low-energy transport in such system can be described using
an effective time-independent Hamiltonian, characterized by
multiple Dirac points in the first Brillouin zone.
Since the quasiparticle spectrum strongly depends on the electromagnetic field, which is polarized along a certain direction, the spectrum around each Dirac point turns out to be anisotropic:
the quasiparticle velocity along the field significantly exceeds the transverse velocity. Moreover, in some of the points the transverse velocity oscillates as a function of the radiation intensity. Such modifications of the spectrum would manifest themselves in quasiparticle transmission through graphene-based junctions.
For instance, we demonstrate that the conductance of a graphene p-n junction depends on the polarization direction as
\begin{eqnarray}
	G\propto |\sin\theta|^{3/2},
	\label{GofTheta}
\end{eqnarray}
where $\theta$ is the angle between the field and the p-n interface, and oscillates as function of the radiation intensity.

Let us emphasize that the previous studies\cite{Syzranov:gapsfirst,Calvo:gaps,Fregoso:topinsgap}
of dynamical gaps generated by linearly polarized radiation addressed {\it the regime of weak irradiation}, $\Delta\ll\omega$, opposite to the limit considered in this paper. At weak radiation the Floquet
spectrum is almost unaltered by the electromagnetic field away from the dynamical gaps. In contrast, we show that in the strong radiation limit there are multiple Dirac points in the first Brillouin zone, and the spectrum is highly anisotropic around each of them.


 {\it The Hamiltonian} of long-wave quasiparticles
in graphene in a single valley subject to external electromagnetic radiation
reads
\begin{eqnarray}
	\cH=v\hbsigma\left[\hat\bp-\frac{e}{c}\bA(t)\right]+U(\br),	
\end{eqnarray}
where $\hat\bp$ is the quasiparticle momentum operator; $\hbsigma$ is the pseudospin operator, the vector of Pauli matrices in the subspace of the two sublattices in graphene, and $U(\br)$ is the electrostatic potential. The electromagnetic field is accounted for by its vector potential $\bA(t)$.

{\it Floquet spectrum.}
 In what immediately follows we derive effective quasiparticle spectra in a uniform potential $U=0$. We choose the coordinates in such a way that $xz$ is the graphene plane, and the  electromagnetic field is directed along the $z$ axis. The gauge potential can be chosen as $A_x=A_y=0$ and $A_z=-cE\omega^{-1}\sin(\omega t)$.
The momentum $\bp$ is then a good quantum number due to the translational invariance of the Hamiltonian, which can be rewritten as
\begin{equation}
	\cH=\hsigma_z\left[ v p_z-\Delta\sin(\omega t)\right]+v\hsigma_x p_x,
	\label{HamReduced}
\end{equation}
where $\Delta=|e|Ev/\omega$.

The condition of large radiation intensity implies $\Delta\gg\omega$. Since the electromagnetic field affects most significantly the
quasiparticles with energies $\varepsilon\lesssim\Delta$,
below we consider sufficiently small momenta, $|v\bp|\ll\Delta$.

\begin{figure}[ht]
	\centering
	\includegraphics[width=0.8\columnwidth]{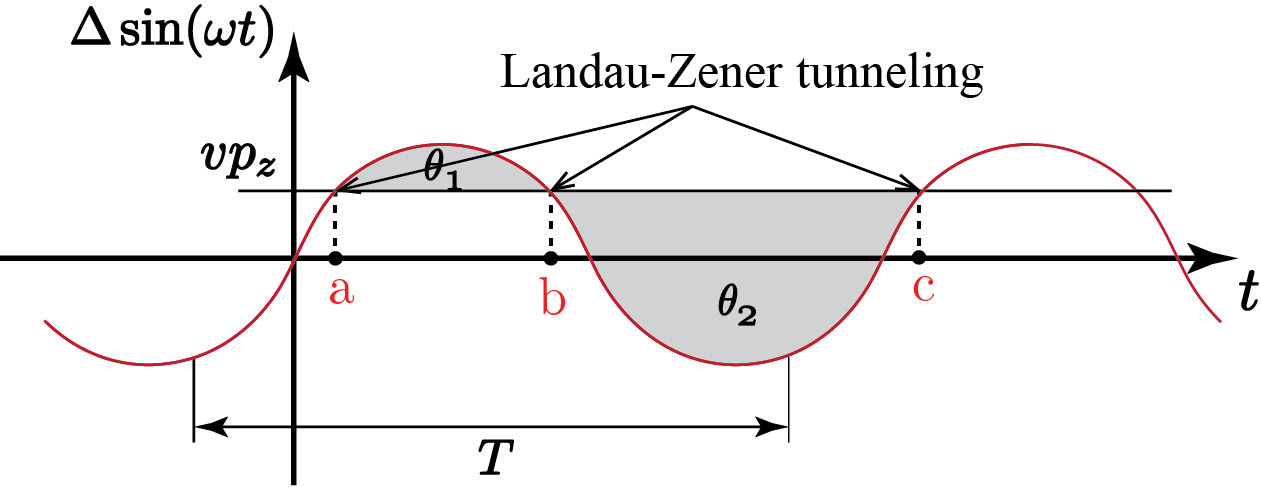}
	\caption{\label{PlotEvol} (Color online) Stages of evolution of the quasiparticle wavefunction. Landau--Zener tunneling between states $\ket{\uparrow}$ and $\ket{\downarrow}$ 	occurs at times a, b, and c. Between these moments the evolution is adiabatic.}
\end{figure}

Except for a close vicinity of the moments $t$ when $\Delta\sin(\omega t)=vp_z$, the second term in the Hamiltonian (\ref{HamReduced}) is small, and the evolution of the quasiparticle wavefunction is adiabatic. The respective diabatic states are close
to the eigenstates of the operator $\hsigma_z$: $\ket{\uparrow}$, an electron moving along the $z$ axis and $\ket{\downarrow}$, an electron moving antiparallel to the $z$ axis. The Hamiltonian (\ref{HamReduced})
is characterized by an avoided-level crossing at $\Delta\sin(\omega t)=vp_z$. At the respective moments of time, Landau--Zener tunneling occurs with probability
\begin{equation}
	P=\exp\left[-\pi v^2p_x^2(\Delta\omega)^{-1}\right].
\end{equation}

Thus, the evolution of a quasiparticle wavefunction on one period $T=2\pi/\omega$ of the oscillating electromagnetic field consists of two intervals of adiabatic evolution, 1) at $\Delta\sin(\omega t)>vp_z$ and 2) at $\Delta\sin(\omega t)<vp_z$, interrupted by
Landau-Zener transitions when $\Delta\sin(\omega t)\approx vp_z$, Fig.~\ref{PlotEvol}. For a given momentum $\bp$, such regime of evolution in the pseudospin space describes the so-called Landau--Zener interferometry~\cite{Shytov:LZinterf,Ashab:LZinterf,Shevchenko:LZinterfreview},
observed previously in superconducting qubits~\cite{Silanpaa:LZinterf,Wilson:LZinterf,Izmalkov:LZinterf,Oliver:LZinterf,Ashab:LZinterf}.


For a given momentum $\bp$, the transformation of the wavefunctions as a result of the adiabatic evolution is described by the transfer matrices
\begin{eqnarray}
	\hat\cR_{1,2}(\bp)=\left(
	\begin{array}{cc}
		e^{-i\theta_{1,2}} & 0 \\
		0 & e^{i\theta_{1,2}}
	\end{array}
	\right)
	\label{R12}
\end{eqnarray}
for intervals 1) and 2), respectively, in the basis $\ket{\uparrow}$ and $\ket{\downarrow}$ in the pseudospin space. In Eq.~(\ref{R12}) we have introduced the dynamical phases accumulated on the respective time intervals:
\begin{subequations}
\begin{equation}
	 \theta_{1,2}(\bp)=\pm\int_{\pm\beta\omega^{-1}}^{(\pi\mp\beta)\omega^{-1}}
	\left\{
	\left[\Delta\sin(\omega t)-vp_z\right]^2+p_x^2v^2
	\right\}^\frac{1}{2}dt,
	\label{thetas}
\end{equation}
\begin{equation}
	\beta=\arcsin(p_zv/\Delta).
	\label{startangle}
\end{equation}
\end{subequations}
The exact values of the dynamical phases are calculated in the Supplemental Material.

The transfer matrices which describe Landau-Zener transitions in the end of intervals ${1}$) and ${2}$)
read respectively
\begin{eqnarray}
	\hat\cL_{1,2}(\bp)=\left(
	\begin{array}{cc}
		\sqrt{P} & \mp\sqrt{1-P}e^{-i\phi_S} \\
		\pm\sqrt{1-P}e^{i\phi_S} & \sqrt{P}
	\end{array}
	\right)
	\label{L12},
\end{eqnarray}
where $\phi_S=\pi/4-p_x^2v^2(2\Delta\omega)^{-1}\ln[p_x^2v^2(2e\Delta\omega)^{-1}]
+\arg\Gamma[ip_x^2v^2(2\Delta\omega)^{-1}]$ is the so-called Stokes phase~\cite{Kayanuma:LZ}.


The evolution of the quasiparticle wavefunction on the period $T=2\pi/\omega$ of the oscillating field is determined by the operator
\begin{equation}
	 \hat\cW(\bp)=\hat\cR_1(\bp)\hat\cL_1(\bp)\hat\cR_2(\bp)\hat\cL_2(\bp).
	\label{EvolutionOperator}
\end{equation}
According to the Floquet theorem~\cite{Hanggi:floquet},
the general solution of a Schr{\"o}dinger equation $i\partial_t\Psi=\cH(t)\Psi$ with a $T$-periodic Hamiltonian can be represented in the form $\Psi(t)=e^{-i\varepsilon t}\Phi(t)$,
where $\Phi(t)$ is a $T$-periodic function, and $\varepsilon$ is the so-called Floquet energy defined up to an integer of $2\pi/T=\omega$.

\begin{figure}[t]
	\centering
	\includegraphics[width=0.73\columnwidth]{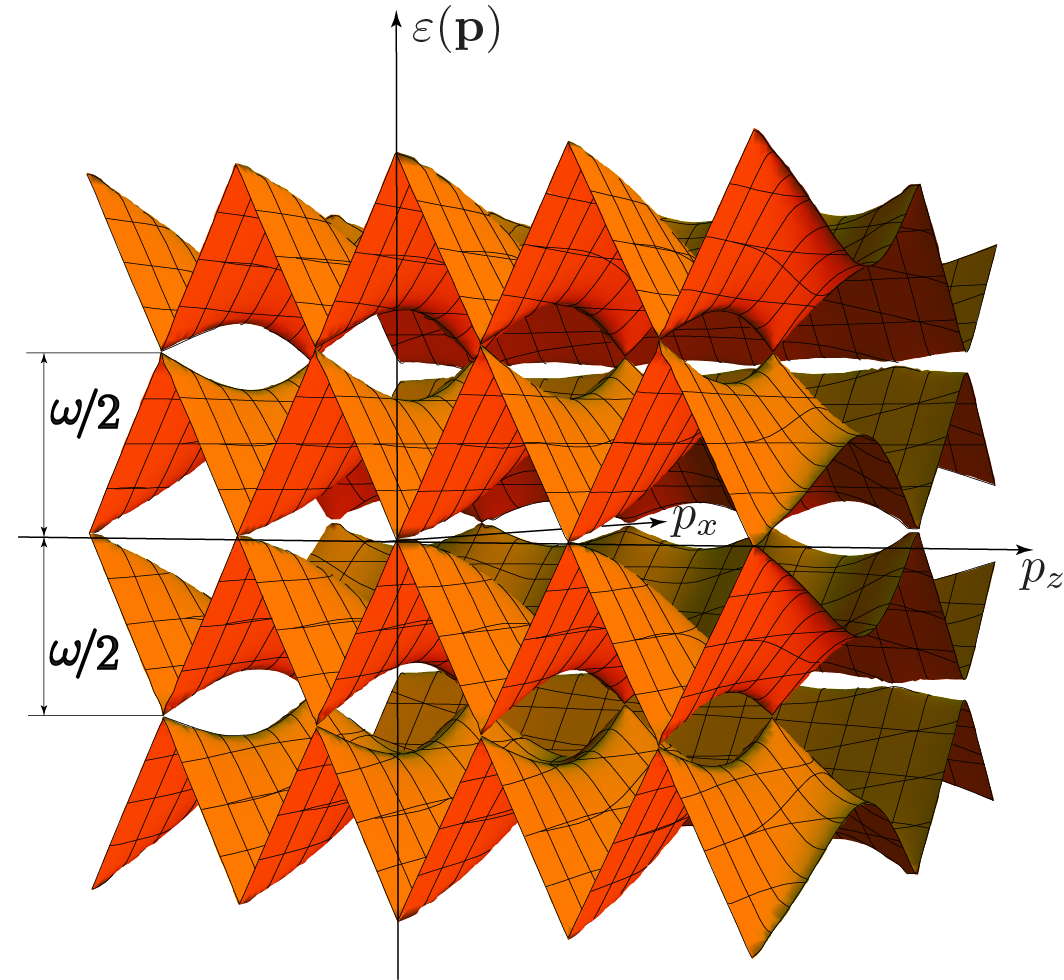}
	\caption{
	\label{Spectra}
	(Color online) Floquet spectra of quasiparticles in strongly irradiated graphene.
	}
\end{figure}

From Eqs.~(\ref{R12}), (\ref{L12}), and (\ref{EvolutionOperator}) we find for the Floquet spectra $\varepsilon_\bp$
of Hamiltonian (\ref{HamReduced})
\begin{equation}
	\cos(\varepsilon_\bp T)=P\cos(\theta_1+\theta_2)+(1-P)\cos(\theta_2-\theta_1).
	\label{SpectraEq}
\end{equation}
The spectra $\varepsilon_\bp$ are plotted in Fig.~\ref{Spectra}. As we show below, the role of the Floquet spectra for transport properties of an irradiated sample are rather similar to those of the usual quasiparticle spectra in a conductor not subject to any time-dependent perturbations.

\begin{figure}[t!]
	\centering
	\includegraphics[width=0.85\columnwidth]{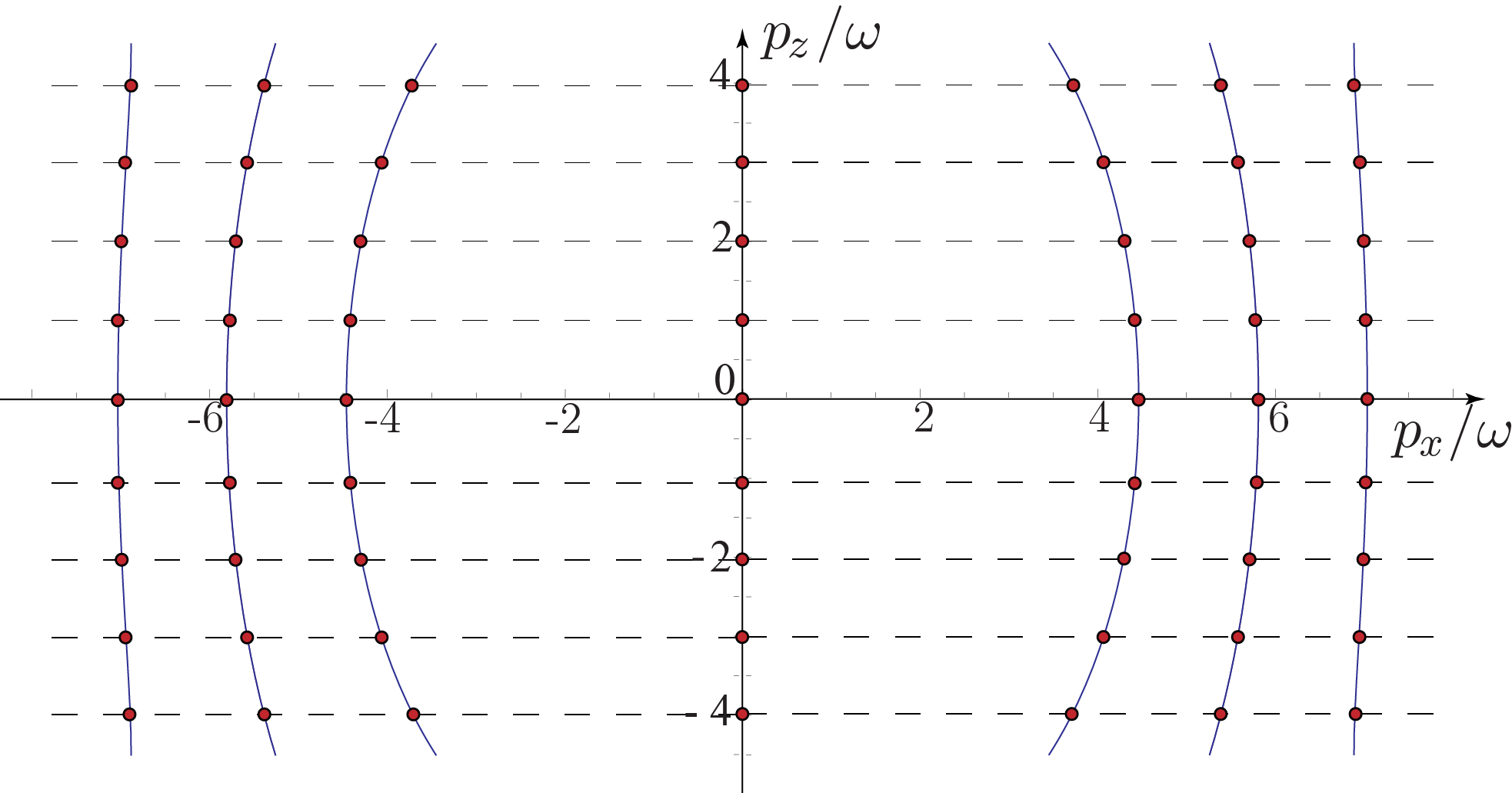}
	\caption{
	\label{Points}
	(Color online) The locations of the Dirac points in the Floquet spectrum.}
\end{figure}

{\it Anisotropic Dirac spectrum.} The obtained Floquet spectrum has multiple Dirac points, Fig.~\ref{Points} (at $\varepsilon_\bp=0$), separated from one another by the characteristic momentum scale $\omega/v$ along the $z$ axis, and $\sqrt{\omega\Delta/v}$ -- along the $x$ axis. For low frequencies $\omega$, considered in this paper,
the number of Dirac points in the first Brillouin zone is large.
There are two types of such points: at $p_x=0$ and at $p_x\neq0$.

In the former case $P=1$, and the locations of the Dirac points are determined by the condition $\theta_1+\theta_2\equiv-2\pi vp_z/\omega=2\pi n$; $n=0,1,2,\ldots$. From Eqs.~(\ref{SpectraEq})
and (\ref{thetas}) we find the anisotropic Dirac spectrum close to the $n$-th Dirac point at small momenta
$|\bp|\ll\Delta/v$:
\begin{subequations}
\begin{eqnarray}
	 \varepsilon_{\bp,n}=\pm\left[(vp_z-n\omega)^2+
v_x^2p_x^2\right]^\frac{1}{2},\\
	 v_x\approx{\omega}({\pi\Delta})^{-1}\left|
\sin({2\Delta}/{\omega})\right|.
	\label{transversevelocity}
\end{eqnarray}
\end{subequations}
Thus, quasiparticles have a strongly anisotropic spectrum near the Dirac points under consideration; the velocity $v_x$ perpendicular to the field is significantly exceeded by the longitudinal velocity $v$
and oscillates as a function of the radiation intensity $S=c\omega^2(8\pi e^2v^2)^{-1}\Delta^2$.

For $p_x\neq0$, the locations of the Dirac points are determined by the conditions $\theta_1\pm\theta_2 =0$ modulo ${2\pi}$. From Eqs.~(\ref{thetas}), (\ref{startangle}), and (\ref{SpectraEq}) we find the anisotropic Dirac spectrum near a Dirac point
located at momenta $|p_z|\ll\Delta/v$ and $0<|p_x|\ll\Delta/v$:
\begin{subequations}
\begin{eqnarray}
	\varepsilon_\bp^2=(v_{zz}\delta p_z+v_{zx}\delta p_x)^2+(v_{xz}\delta p_z+v_{xx}\delta p_x)^2,
	\label{SpectrumNonTriv}\\
	v_{zz}=Pv,
		\label{vzz} \\
	v_{zx}=-2(1-P)p_xv(\pi\Delta)^{-1}\log(\Delta/|vp_x|),\\
	 v_{xx}=-2[P(1-P)]^\frac{1}{2}p_xv(\pi\Delta)^{-1}\log(\Delta/|vp_x|),\\
	v_{xz}=-[P(1-P)]^\frac{1}{2}v	
	\label{vxz},
\end{eqnarray}
\end{subequations}
where $\delta p_x$ and $\delta p_z$ are the deviations of the momenta $p_x$ and $p_z$ from their values at the Dirac point.
The spectrum (\ref{SpectrumNonTriv}) is also anisotropic; the velocities $v_{zz}$ and $v_{xz}$, which characterize motion along the $z$ axis, significantly exceed the velocities $v_{xx}$ and $v_{zx}$, characterizing the transverse motion. Unlike the case of $p_x=0$, the velocities do not oscillate as a function of the radiation intensity.

{\it Effective Hamiltonian.} Let us demonstrate that
quasiparticle dynamics close to the Dirac points can be described
by an effective time-independent Dirac Hamiltonian.
Indeed,
on one period $T$, the evolution operator of a quasiparticle wavefunction can be represented as
\begin{equation}
	\hat\cW_{T}(\bp)= e^{-iT\cH_{\rm eff}(\bp)},
\end{equation}
where the eigenvalues of the effective time-independent Hamiltonian $\cH_{\rm eff}(\bp)$ are given (up to an integer number of $\omega$) by the Floquet spectrum $\varepsilon_\bp$.

To describe quasiparticle transport in a smooth coordinate-dependent potential $U(\br)$ it is sufficient to consider the wavepacket evolution in discrete time with interval $T$. Due to the smallness of the gradient of the potential, the momentum change on one period is small, and the evolution is equivalent to that for a time-independent problem with Hamiltonian $\cH_{\rm eff}(\bp)+U(\br)$. A detailed justification of the effective Hamiltonian method is presented in the Supplemental Material.

In principle, such approach can be used to analyze transport of quasiparticles with an arbitrary effective Hamiltonian $\cH_{\rm eff}$ of the kinetic energy, provided the potential is sufficiently smooth. In this paper, we consider
quasiparticle dynamics close to the Dirac points, so that the eigenvalues of $\cH_{\rm eff}(\bp)$ can be chosen small, $|\varepsilon_\bp|\ll\omega$, corresponding to a small change of quasiparticle wavefunctions during one period, $\hat\cW_{T}(\bp)\approx {1-iT\cH_{\rm eff}(\bp)}$.
Near the Dirac points with $p_x=0$ we find from Eqs.~(\ref{R12})-(\ref{EvolutionOperator})
\begin{equation}
	\cH_{\rm eff}(\bp)=v\delta p_z\hsigma_z+v_x\delta p_x\hsigma_x,
	\label{Htrivial}
\end{equation}
where $\delta p_x$ and $\delta p_z$ are the momentum deviations from the Dirac points along the axes $x$ and $z$, respectively, and the renormalized transverse velocity $v_x$ is defined by Eq.~(\ref{transversevelocity}).

Near the other Dirac points, at $p_x\neq0$,
\begin{equation}
	\cH_{\rm eff}(\bp)=(v_{zz}\delta p_z+v_{zx}\delta p_x)\hsigma_z+(v_{xx}\delta p_x+v_{xz}\delta p_z)\hsigma_x,
	\label{Hnontrivial}
\end{equation}
where the velocities $v_{zz}$, $v_{zx}$,  $v_{xz}$, and $v_{xx}$ are defined by Eqs.~(\ref{vzz})-(\ref{vxz}).

{\it Transmission through potential barriers.}
In a smooth coordinate-dependent potential $U(\br)$ the quasiparticle motion is quasiclassical, except for a small vicinity of the classical turning points, where $\varepsilon=U(\br)$. To find the transmission coefficient near such turning point one can approximate the potential by a linear function $U(\br)\approx \varepsilon+\bF\br$, where $\bF$ is the gradient of the potential.

The component $p_\bot$ of the momentum, perpendicular to $\bF$, is conserved. Representing the Hamiltonian near a Dirac point as
$\cH_{\rm eff}=(\bv_\parallel\boldsymbol\sigma)\delta p_\parallel
+(\bv_\bot\boldsymbol\sigma)\delta p_\bot$ and
using the coordinate operator $\br=i\partial_{\bf \delta p}$, we arrive at the Schr\"odinger equation for the quasiparticle motion near the turning point in the form
\begin{eqnarray}
	\left[(\bv_\parallel\boldsymbol\sigma)\delta p_\parallel+
	(\bv_\bot\boldsymbol\sigma)\delta p_\bot+iF\partial_{\delta p_\parallel}\right]\Psi=0.
	\label{LZSchroed}
\end{eqnarray}
Eq.~(\ref{LZSchroed}) describes Landau--Zener tunnelling
in momentum space, similarly to that for the usual isotropic Dirac spectrum in graphene\cite{CheianovFalko,Shytov:LZmagn}.

\begin{figure}[t!]
	\centering
	\includegraphics[width=0.75\columnwidth]{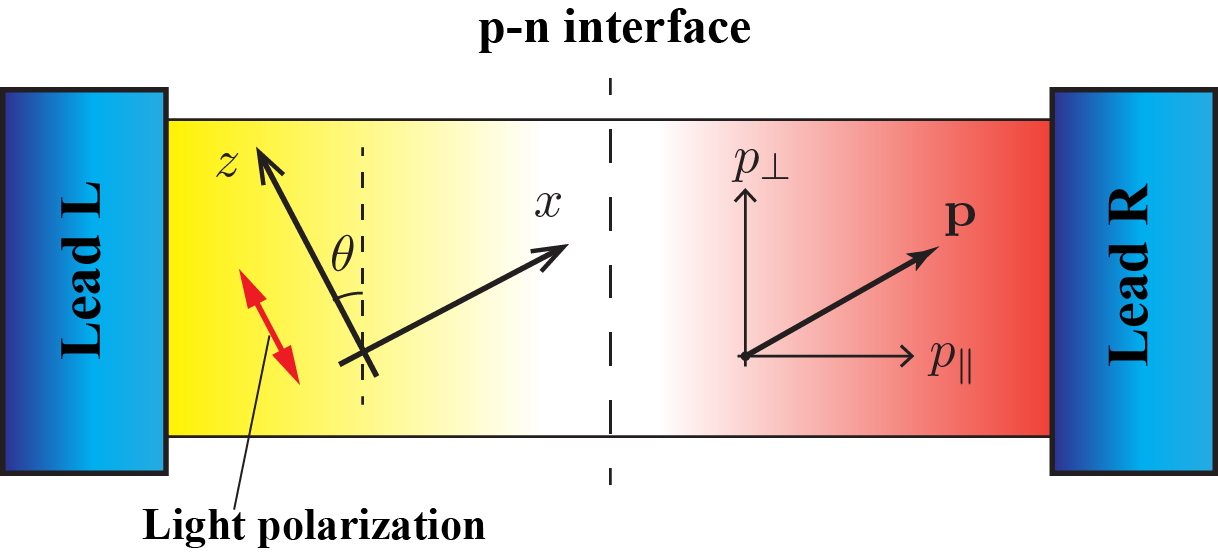}
	\caption{\label{Strip} (Color online) Irradiated graphene junction.}
\end{figure}

After some algebra (cf. Supplemental Material for details), we find the transmission coefficient of a quasiparticle with an anisotropic
Dirac Hamiltonian (\ref{Hnontrivial}) through a potential barrier with slope $F$
\begin{eqnarray}
	T(\delta p_\bot)=
	\nonumber\\
	\exp
	\left[-\pi\: \delta p_\bot^2|v_{xx}v_{zz}-v_{xz}v_{zx}|^2/
	\left(FP^{3/2}|v\sin\theta|^{3}\right)\right],
	\label{Transmission}
\end{eqnarray}
where $\theta$ is the angle between the light polarization and the transverse direction, Fig.~(\ref{Strip}). The transverse momentum $\delta p_\bot$ is counted from the Dirac point.

The transmission coefficient, Eq.~(\ref{Transmission}), describes quasiparticle penetration through the potential barrier near the Dirac points at $p_x=0$, as well as those at $p_x\neq0$, because the effective Hamiltonian in both cases has the form of Eq.~(\ref{Hnontrivial}), which we used to derive Eq.~(\ref{Transmission}).
The case of $p_x=0$ corresponds to $v_{zz}=v$, $v_{xx}=v_x$, and $v_{xz}=v_{xz}=0$, cf. Eq.~(\ref{transversevelocity}).

Quasiparticles penetrate through a potential barrier without reflection in the case of a normal incidence (Klein paradox); $T(\delta p_\bot=0)=1$, similarly to isotropic-spectrum Dirac fermions~\cite{CheianovFalko}. The transmission coefficient decreases with increasing $\delta p_\bot$ and with decreasing the angle $\theta$, $\log T\propto-\delta p_\bot^2|\sin\theta|^{-3}$.

Eq.~(\ref{Transmission}) applies for angles $\theta$, which are not too small; $\theta\gtrsim\min(\omega/\Delta,pv/\Delta)$. The quasiparticle velocity is strongly suppressed in the direction perpendicular to the electromagnetic field, so if its
polarization is perpendicular to the potential barrier, $\theta\rightarrow0$, quasiparticles move towards the barrier with very low speeds, and the transmission is suppressed, $T\rightarrow0$.

{\it Conductance.}
Let us consider the conductance of a strongly irradiated ballistic p-n junction, Fig.~\ref{Strip}. We assume that the potential is uniform far from the the p-n interface and that the angle
between the electromagnetic field ($z$ axis) and the interface is $\theta$.

In the framework of the effective-Hamiltonian description,
the statistics of electrons sufficiently far on the left and on the right from the interface can be characterised by distribution functions $f_L[\varepsilon(\bp)]$ and $f_R[\varepsilon(\bp)]$, which coincide
with each other in the absence of an external bias voltage applied to the junction. The exact form of each distribution function
depends on the details of the relaxation and the interaction of quasiparticles with radiation far from the interface or in the electrodes. As we show below, the conductance does not depend on the form of the distribution functions; we require only that $f_{R,L}$
vanish at sufficiently high energies and saturate to unity at sufficiently low energies.

In fact, $f_L[\varepsilon(\bp)]$ and $f_R[\varepsilon(\bp)]$ are given by the Fermi distribution function $f_F[\varepsilon(\bp)]$ if
electrons far from the interface are being equilibrated by a bath of excitations, e.g., phonons, with a temperature sufficiently smaller than the frequency $\omega$, which corresponds to realistic
temperatures and to the required frequencies, as we estimate below.
Indeed, in that case the phonon degrees of freedom are slow compared to the electromagnetic field and can be included in the effective
Hamiltonian, which would correspond then to a non-driven electronic system equilibrated by a phonon bath.

The current through the junction reads~\cite{Syzranov:gapsfirst}
\begin{equation}
	I=2eW\int\frac{d\bp}{(2\pi)^2}\frac{\partial\varepsilon}{\partial p_\parallel} T(p_\bot)
	\left\{f_L[\varepsilon(\bp)]-f_R[\varepsilon(\bp)]\right\},
	\label{I}
\end{equation}
where 
$W$ is the width of the graphene strip, $\partial\varepsilon/\partial p_\parallel$ is the longitudinal velocity far from the interface.

The transverse momentum $p_\bot$ is conserved during the motion and determines the transmission coefficient through the junction, Eq.~(\ref{Transmission}). Quasiparticles penetrate through the potential barrier only if their transverse momenta are
sufficiently close to those at the Dirac points. In a smooth potential barrier only the quasiparticles from narrow intervals $\propto F^{1/2}$ of the transverse momenta penetrate efficiently through the barrier, allowing for an independent integration with respect to the transverse momentum $p_\bot$ in Eq.~(\ref{I}) around each Dirac point
\begin{equation}
	\int T(\delta p_\bot) d\delta p_\bot\propto|\sin\theta|^\frac{3}{2},
\end{equation}
which describes the dependence of the conductance $G(\theta)$ on the angle $\theta$, Eq.~(\ref{GofTheta}).

{\it Conductance oscillations.}
In general, each Dirac point for $p\lesssim\Delta/v$ contributes to the conductance of the junction. Provided the momentum $\bp$ at a Dirac point is sufficiently smaller than the characteristic
quasiparticle momenta away from the interface, the contribution of this point to the conductance can be found straightforwardly from Eq.~(\ref{I}), considering that in the presence of a small voltage $V$, applied to the junction, $f_L(\varepsilon)=f_R(\varepsilon-eV)$:
\begin{eqnarray}
	G_{\rm one Dirac \atop point}=
	 \frac{We^2}{2\pi^2}\frac{F^\frac{1}{2}P^\frac{3}{4}v^\frac{3}{2}
|\sin\theta|^\frac{3}{2}}
	{|v_{xx}v_{zz}-v_{xz}v_{zx}|}.
\end{eqnarray}

While for the Dirac points at $p_x\neq 0$ the conductance is a smooth function of the radiation intensity, the conductance due to the points at $p_x=0$ oscillates as a function of the radiation intensity $S$
\begin{equation}
 G_0=\frac{2^\frac{1}{2}|e|^3WF^\frac{1}{2}S^\frac{1}{2}
 |v\sin\theta|^\frac{3}{2}}{\pi^\frac{1}{2}\omega^2c^\frac{1}{2}}
	\left|\sin\!\!\left(
	 \frac{4(2\pi)^\frac{1}{2}|e|vS^\frac{1}{2}}{\omega^2c^\frac{1}{2}}
	\right)\right|^{-1}.
\end{equation}
As a result, the entire conductance of the junction oscillates with the radiation intensity.

{\it Estimates.} Let us summarize the conditions necessary for the observation of the discussed effects. To ensure quantum coherence on one period of the electromagnetic field and the smoothness of
the potential, the period has to be smaller than the (elastic or inelastic) scattering time $\tau$ and the characteristic time of the traversal of the
potential barrier in the graphene junction.

For a ballistic 300 nm junction, we obtain $\omega\gtrsim$ 20 THz.
The regime of strong radiation intensity, $\Delta\gg\omega$,
requires
\begin{equation}
	 S\gtrsim\frac{c}{8\pi}\left(\frac{\hbar\omega^2}
{|e|v}\right)^2\approx40\frac{kW}{cm^2}.
\end{equation}
The required radiation intensity grows rapidly with the frequency and decreases with the junction length $L$, $\propto\omega^4,L^{-4}$.

The period of conductance oscillations is of the order of the minimal required radiation intensity. These oscillations, being a consequence of single-particle interference on the period of the electromagnetic
field, do not involve interference of different quasiparticle trajectories, unlike, e.g., the effects studied in Refs.~\onlinecite{Fistul:interference} and \onlinecite{YampNori:oscilcond1}, and are insensitive to quenched disorder under the above specified conditions.


{\it Acknowledgements.} We acknowledge useful discussions with U.~Briskot, I.~Burmistrov, A.V.~Shytov, and A.G.~Semenov.
This work was partially supported by the ARO, RIKEN iTHES Project,
MURI Center for Dynamic Magneto-Optics, JSPS-RFBR contract no. 12-02-92100, Grant-in-Aid for Scientific Research (S), MEXT Kakenhi on Quantum Cybernetics, the JSPS via its FIRST program, and by the Russian Foundation for Basic Research (projects 11-02-00708, 11-02-00741, and 12-02-00339). Ya.I.R. acknowledges support from
the Federal Program `Human Capital for Science and
Education in Innovative Russia' (grants 8169 and 8171) and the
National University of Science and Technology `MISIS' (grant No.~3400022).

\newpage
\clearpage

\section*{Supplemental material\label{Sec:App}}
\numberwithin{equation}{subsection}
\subsection{Dynamical phases}
In this Section we present a detailed computation of the dynamical phases $\theta_{1,2}$ in the limit $|\bp|\ll\Delta/v$.
For $p_z=0$
\begin{gather}
  \begin{split}
  \theta_{1,2}&=\pm\frac{\Delta}{\omega}\int\limits_{0}^\pi \big\{\sin^2\vf+(p_xv/\Delta)^2\big\}^{1/2}d\varphi\\
  &=\pm\frac{2p_xv}{\omega}E\left[-\frac{\Delta^2}{(vp_x)^2}\right]
  \\
  &\approx\pm\frac{\Delta}{\omega}\left(2+\frac{|vp_x|}
  {\Delta}\ln\frac{\Delta}{|vp_x|}\right),\ \ vp_x\ll\Delta,
  \label{Elliptic}
 \end{split}
\end{gather}
$E(x)$ being the complete elliptic integral of the second kind. In the last line of Eq.~(\ref{Elliptic}), we used the $x\gg1$ asymptotic of $E(x)$.

Let us find phase $\theta_1$ for $p_z\neq0$ and $\Delta\gg vp$.
Introducing new variables $\sin\varphi=t,\ a=\arcsin(p_z/\Delta),\ b=p_x/\Delta$, we expand the phase
\begin{gather}
  \begin{split}
   \theta_1&=\frac{\Delta}{\omega}\int\limits_{\beta}^{\pi-\beta}
   \left[\left(\sin\varphi-\frac{vp_z}{\Delta}\right)^2+
   \Big(\frac{vp_x}{\Delta}\Big)^2\right]^{1/2}d\varphi
   \\
   &=\frac{2\Delta}{\omega}\int\limits_{a}^1\frac{\sqrt{(t-a)^2+b^2}}
   {\sqrt{1-t^2}}dt=\frac{2\Delta}{\omega}I(a,b)
  \end{split}
\end{gather}
in terms of small $b$:
\begin{eqnarray}
  I(a,b)=C+D b^2\ln b^2+...,	
\end{eqnarray}
\begin{gather}
  \begin{split}
   \frac{\partial I(a,b)}{\partial b^2}& =D \ln b^2e+...=\int\limits_{a}^1\frac{dt}{\sqrt{1-t^2}\sqrt{(t-a)^2+b^2}}=\\
  &\int\limits_{a}^1\frac{dt}{\sqrt{(t-a)^2+b^2}}\Big[\frac{1}
  {\sqrt{1-t^2}}-\frac{1}{\sqrt{1-a^2}}\Big]\\
  &+\frac{1}{\sqrt{1-a^2}}\ln\frac{2(1-a)}{b}+....
  \label{DymPhasA}
  \end{split}
\end{gather}
The first integral in Eq.~(\ref{DymPhasA}) can be taken by setting $b=0$ in it, which yields
\begin{gather}
   D \ln b^2e=\frac{1}{\sqrt{1-a^2}}\ln\frac{4(1-a^2)}{b}.
\end{gather}
Thus, in the limit $vp_z/\Delta\ll1$,  we obtain
\begin{gather}
  \label{phases1}
  \begin{split}
   \theta_1&=\frac{\Delta}{\omega}\bigg[2+
   \Big(\frac{vp_z}{\Delta}\Big)^2-\frac{\pi v p_z}{\Delta}+\frac{v^2p_x^2}{\Delta^2}
\ln\frac{4\sqrt{e}\Delta}{|vp_x|}+\Bigg]...\\
  \end{split}
\end{gather}
The second dynamical phase $\theta_2$ can be found analogously,
\begin{gather}
  \label{phases2}
  \begin{split}
   \theta_2&=-\frac{\Delta}{\omega}\bigg[2+\Big(\frac{vp_z}
   {\Delta}\Big)^2+\frac{\pi v p_z}{\Delta}+\frac{v^2p_x^2}{\Delta^2}
   \ln\frac{4\sqrt{e}\Delta}{|vp_x|}+\Bigg]...
  \end{split}
\end{gather}
Using Eqs.~\eqref{phases1}-\eqref{phases2} we recover the quasiparticle spectrum, Eqs.~(11a)-(11e).

\subsection{Effective-Hamiltonian approach}
In this Section, we present a detailed rigorous justification of using an effective time-independent
Hamiltonian $\cH_{\rm eff}(\bp)$ for the description of quasiparticle dynamics in a coordinate-dependent potential in the presence of
a time-periodic electromagnetic field.

The low-energy effective Hamiltonian is defined through the evolution operator on one period $T$ of the electromagnetic field in a uniform potential ($U=0$),
\begin{eqnarray}
	\hat\cW_T(\bp)
	\equiv T_t e^{-i\int_0^T\cH(t,\bp)dt}
	\nonumber\\	
	=e^{-iT\cH_{\rm eff}(\bp)}\approx1-iT\cH_{\rm eff}(\bp),
\end{eqnarray}
where $T_t$ is our convention for the chronological ordering.

Let us consider quasiparticle motion in a smooth non-uniform potential, which changes insignificantly on the characteristic path $v/\omega$ of a quasiparticle on one period. On such distance the potential can be considered linear and can be taken into account by including a small perturbation $\hV(\br)=\bF\br\equiv i\bF\partial_\bp$ in the evolution operator:

\begin{eqnarray}
	\hat\cW_T^\prime(\bp)=T_t e^{-i\int_0^T(\cH+\hV)dt}
	\approx1-iT\cH_{\rm eff}(\bp)
	\nonumber\\
	-i\int_0^T\hat\cW_t^\dagger(\bp)\hV\hat\cW_t(\bp)dt
	\\
	=1-iT\cH_{\rm eff}-iT\hV
	\nonumber\\
	+F\int_0^T\hat\cW_t^\dagger(\bp)\partial_\bp\hat\cW_t(\bp)dt.
	\label{newW}
\end{eqnarray}
The smoothness of the potential implies the smallness of the perturbation $\hV$ and requires
\begin{eqnarray}
	F\ll\omega^2/v.
\end{eqnarray}

In what immediately follows, we demonstrate that the last term in Eq.~(\ref{newW}) is not important for the quasiparticle dynamics close to Dirac points and, thus, can be disregarded.

First of all, let us notice that the momentum dependence of this term can be neglected.
Indeed, small deviations $\delta p\ll \omega/v$ of momenta from the Dirac point lead to only small modifications $\ll\omega$ of the matrix elements of the Hamiltonian $\cH(t)$, which result in small relative corrections to the last line of Eq.~(\ref{newW}).

Thus, the evolution operator takes the form
\begin{eqnarray}
	\hat\cW_T^\prime(\bp)=1-iT\cH_{\rm eff}-iT\hV-iTF(\hbsigma\bA+A_0),
\end{eqnarray}
where $\bA$ and $A_0$ are momentum-independent, $|\bA|, A_0\lesssim v$, and, thus, are of no consequence for any physical observables.

To conclude, on the interval $T$ the evolution operator for quasiparticle dynamics in a non-uniform potential $\hV$ corresponds to the time-independent Hamiltonian
\begin{eqnarray}
	\cH_{\rm eff}^\prime=\cH_{\rm eff}+\hV,
\end{eqnarray}
which justifies the applicability of the effective-Hamiltonian approach.

\subsection{Transmission coefficient}
In this Section we consider the transmission of anisotropic Dirac quasiparticles
through a potential barrier.
The effective Hamiltonian, Eq.~(\ref{Hnontrivial}), can be conveniently represented in the form
\begin{gather}
   {\cH}_{\rm eff}=\bm{\sigma}\mathbf{V}\delta\mathbf{p},\ \ \mathbf{V}=\begin{pmatrix}
                  v_{xx}\ &\ v_{xz}\\
                  v_{zx}\ & v_{zz}
   \end{pmatrix}
\end{gather}
Performing a basis rotation in the momentum space,
 \begin{gather}
\delta \tilde{\mathbf{p}}=
\begin{pmatrix}
   \delta p_\parallel\\
   \delta p_\perp
\end{pmatrix}=
\begin{pmatrix}
   \cos\theta & -\sin\theta\\
   \sin\theta &  \cos\theta
\end{pmatrix}\begin{pmatrix}
   \delta p_x\\
   \delta p_z
\end{pmatrix}\equiv\mathbf{S}\delta \mathbf{p},
 \end{gather}
we arrive at
\begin{gather}
   \big[\bm{\sigma}\mathbf{VS}^{-1}\delta \tilde{\mathbf{p}}+iF\de_{\delta p_\parallel}\big]\Psi=0,
\end{gather}
which describes Landau--Zener tunneling in the momentum space.
Its probability can be recast in basis-invariant form as
 \begin{gather}
   \label{transm}
   T(p_z)=\exp\Big[-\frac{\pi\delta p_\perp^2|\hbox{det}\mathbf{V}|^2}{F|\mathbf{v}_{||}|^3}\Big]
\end{gather}
where $\mathbf{v}_\parallel=([\mathbf{VS}^{-1}]_{xx},[\mathbf{VS}^{-1}]_{zx})$.
To evaluate $v_\parallel$ we neglect the velocities $v_{zz}$ and $v_{zx}$ in comparison with
$v_{xx}$ and $v_{xz}$,
\begin{gather}
  \mathbf{v}^2\approx (v_{xz}^2+v_{zz}^2)\sin^2\theta=Pv^2\sin^2\theta,
\end{gather}
which, combined with Eq.~\eqref{transm}, leads to Eq.~(16).


\begin{thebibliography}{22}
\expandafter\ifx\csname natexlab\endcsname\relax\def\natexlab#1{#1}\fi
\expandafter\ifx\csname bibnamefont\endcsname\relax
  \def\bibnamefont#1{#1}\fi
\expandafter\ifx\csname bibfnamefont\endcsname\relax
  \def\bibfnamefont#1{#1}\fi
\expandafter\ifx\csname citenamefont\endcsname\relax
  \def\citenamefont#1{#1}\fi
\expandafter\ifx\csname url\endcsname\relax
  \def\url#1{\texttt{#1}}\fi
\expandafter\ifx\csname urlprefix\endcsname\relax\def\urlprefix{URL }\fi
\providecommand{\bibinfo}[2]{#2}
\providecommand{\eprint}[2][]{\url{#2}}

\bibitem[{\citenamefont{Galitskii et~al.}(1970)\citenamefont{Galitskii,
  Goreslavskii, and Elesin}}]{Galitskii:firstgap}
\bibinfo{author}{\bibfnamefont{V.~M.} \bibnamefont{Galitskii}},
  \bibinfo{author}{\bibfnamefont{S.~P.} \bibnamefont{Goreslavskii}},
  \bibnamefont{and} \bibinfo{author}{\bibfnamefont{V.~F.}
  \bibnamefont{Elesin}}, \bibinfo{journal}{Sov. Phys. JETP}
  \textbf{\bibinfo{volume}{30}}, \bibinfo{pages}{117} (\bibinfo{year}{1970}).

\bibitem[{\citenamefont{Galitskii and
  Elesin}(1969)}]{GoreslavskiiElesin:metalinsulator}
\bibinfo{author}{\bibfnamefont{V.~M.} \bibnamefont{Galitskii}}
  \bibnamefont{and} \bibinfo{author}{\bibfnamefont{V.~F.}
  \bibnamefont{Elesin}}, \bibinfo{journal}{JETP Lett.}
  \textbf{\bibinfo{volume}{10}}, \bibinfo{pages}{491} (\bibinfo{year}{1969}).

\bibitem[{\citenamefont{Alexandrov and
  Elesin}(1972)}]{AlexandrovElesin:absorption}
\bibinfo{author}{\bibfnamefont{A.~S.} \bibnamefont{Alexandrov}}
  \bibnamefont{and} \bibinfo{author}{\bibfnamefont{V.~F.}
  \bibnamefont{Elesin}}, \bibinfo{journal}{Sov. Phys. JETP}
  \textbf{\bibinfo{volume}{35}}, \bibinfo{pages}{403} (\bibinfo{year}{1972}).

\bibitem[{\citenamefont{Fistul et~al.}(2010)\citenamefont{Fistul, Syzranov,
  Kadigrobov, and Efetov}}]{Fistul:interference}
\bibinfo{author}{\bibfnamefont{M.~V.} \bibnamefont{Fistul}},
  \bibinfo{author}{\bibfnamefont{S.~V.} \bibnamefont{Syzranov}},
  \bibinfo{author}{\bibfnamefont{A.~M.} \bibnamefont{Kadigrobov}},
  \bibnamefont{and} \bibinfo{author}{\bibfnamefont{K.~B.}
  \bibnamefont{Efetov}}, \bibinfo{journal}{Phys. Rev. B}
  \textbf{\bibinfo{volume}{82}}, \bibinfo{pages}{121409(R)}
  (\bibinfo{year}{2010}).

\bibitem[{\citenamefont{Oka and Aoki}(2009)}]{OkaAokiHall:Hall}
\bibinfo{author}{\bibfnamefont{T.}~\bibnamefont{Oka}} \bibnamefont{and}
  \bibinfo{author}{\bibfnamefont{H.}~\bibnamefont{Aoki}},
  \bibinfo{journal}{Phys. Rev. B} \textbf{\bibinfo{volume}{79}},
  \bibinfo{pages}{081406(R)} (\bibinfo{year}{2009}).

\bibitem[{\citenamefont{Elesin et~al.}(1979)\citenamefont{Elesin, Erko, and
  Larkin}}]{Elesin:firstgapobservation}
\bibinfo{author}{\bibfnamefont{V.~F.} \bibnamefont{Elesin}},
  \bibinfo{author}{\bibfnamefont{A.~I.} \bibnamefont{Erko}}, \bibnamefont{and}
  \bibinfo{author}{\bibfnamefont{A.~I.} \bibnamefont{Larkin}},
  \bibinfo{journal}{JETP Lett.} \textbf{\bibinfo{volume}{29}},
  \bibinfo{pages}{651} (\bibinfo{year}{1979}).

\bibitem[{\citenamefont{Vu et~al.}(2004)\citenamefont{Vu, Haug, M{\"u}cke,
  Tritschler, Wegener, Khitrova, and Gibbs}}]{Vu:Mollowtriplet}
\bibinfo{author}{\bibfnamefont{Q.~T.} \bibnamefont{Vu}},
  \bibinfo{author}{\bibfnamefont{H.}~\bibnamefont{Haug}},
  \bibinfo{author}{\bibfnamefont{O.~D.} \bibnamefont{M{\"u}cke}},
  \bibinfo{author}{\bibfnamefont{T.}~\bibnamefont{Tritschler}},
  \bibinfo{author}{\bibfnamefont{M.}~\bibnamefont{Wegener}},
  \bibinfo{author}{\bibfnamefont{G.}~\bibnamefont{Khitrova}}, \bibnamefont{and}
  \bibinfo{author}{\bibfnamefont{H.~M.} \bibnamefont{Gibbs}},
  \bibinfo{journal}{Phys. Rev. Lett.} \textbf{\bibinfo{volume}{92}},
  \bibinfo{pages}{217403} (\bibinfo{year}{2004}).

\bibitem[{\citenamefont{Syzranov et~al.}(2008)\citenamefont{Syzranov, Fistul,
  and Efetov}}]{Syzranov:gapsfirst}
\bibinfo{author}{\bibfnamefont{S.~V.} \bibnamefont{Syzranov}},
  \bibinfo{author}{\bibfnamefont{M.~V.} \bibnamefont{Fistul}},
  \bibnamefont{and} \bibinfo{author}{\bibfnamefont{K.~B.}
  \bibnamefont{Efetov}}, \bibinfo{journal}{Phys. Rev. B}
  \textbf{\bibinfo{volume}{78}}, \bibinfo{pages}{045407}
  (\bibinfo{year}{2008}).

\bibitem[{\citenamefont{Calvo et~al.}(2011)\citenamefont{Calvo, Pastawski,
  Roche, and Torres}}]{Calvo:gaps}
\bibinfo{author}{\bibfnamefont{H.~L.} \bibnamefont{Calvo}},
  \bibinfo{author}{\bibfnamefont{H.~M.} \bibnamefont{Pastawski}},
  \bibinfo{author}{\bibfnamefont{S.}~\bibnamefont{Roche}}, \bibnamefont{and}
  \bibinfo{author}{\bibfnamefont{L.~E. F.~F.} \bibnamefont{Torres}},
  \bibinfo{journal}{Appl. Phys. Lett.} \textbf{\bibinfo{volume}{98}},
  \bibinfo{pages}{232103} (\bibinfo{year}{2011}).

\bibitem[{\citenamefont{Fregoso et~al.}(2013)\citenamefont{Fregoso, Wang,
  Gedik, and Galitski}}]{Fregoso:topinsgap}
\bibinfo{author}{\bibfnamefont{B.~M.} \bibnamefont{Fregoso}},
  \bibinfo{author}{\bibfnamefont{Y.}~\bibnamefont{Wang}},
  \bibinfo{author}{\bibfnamefont{N.}~\bibnamefont{Gedik}}, \bibnamefont{and}
  \bibinfo{author}{\bibfnamefont{V.}~\bibnamefont{Galitski}}
  (\bibinfo{year}{2013}), \bibinfo{note}{arXiv:1305.4145}.

\bibitem[{\citenamefont{Shytov et~al.}(2003)\citenamefont{Shytov, Ivanov, and
  Feigelman}}]{Shytov:LZinterf}
\bibinfo{author}{\bibfnamefont{A.~V.} \bibnamefont{Shytov}},
  \bibinfo{author}{\bibfnamefont{D.}~\bibnamefont{Ivanov}}, \bibnamefont{and}
  \bibinfo{author}{\bibfnamefont{M.~V.} \bibnamefont{Feigelman}},
  \bibinfo{journal}{Eur. Phys. J. B} \textbf{\bibinfo{volume}{36}},
  \bibinfo{pages}{263} (\bibinfo{year}{2003}).

\bibitem[{\citenamefont{Ashab et~al.}(2007)\citenamefont{Ashab, Johansson,
  Zagoskin, and Nori}}]{Ashab:LZinterf}
\bibinfo{author}{\bibfnamefont{S.}~\bibnamefont{Ashab}},
  \bibinfo{author}{\bibfnamefont{J.}~\bibnamefont{Johansson}},
  \bibinfo{author}{\bibfnamefont{A.}~\bibnamefont{Zagoskin}}, \bibnamefont{and}
  \bibinfo{author}{\bibfnamefont{F.}~\bibnamefont{Nori}},
  \bibinfo{journal}{Phys. Rev. A} \textbf{\bibinfo{volume}{75}},
  \bibinfo{pages}{063414} (\bibinfo{year}{2007}).

\bibitem[{\citenamefont{Shevchenko et~al.}(2010)\citenamefont{Shevchenko,
  Ashhab, and Nori}}]{Shevchenko:LZinterfreview}
\bibinfo{author}{\bibfnamefont{S.}~\bibnamefont{Shevchenko}},
  \bibinfo{author}{\bibfnamefont{S.}~\bibnamefont{Ashhab}}, \bibnamefont{and}
  \bibinfo{author}{\bibfnamefont{F.}~\bibnamefont{Nori}},
  \bibinfo{journal}{Phys. Rep.} \textbf{\bibinfo{volume}{492}},
  \bibinfo{pages}{1} (\bibinfo{year}{2010}).

\bibitem[{\citenamefont{Sillanp{\"a}{\"a}
  et~al.}(2005)\citenamefont{Sillanp{\"a}{\"a}, Lehtinen, Paila, Makhlin, and
  Hakonen}}]{Silanpaa:LZinterf}
\bibinfo{author}{\bibfnamefont{M.}~\bibnamefont{Sillanp{\"a}{\"a}}},
  \bibinfo{author}{\bibfnamefont{T.}~\bibnamefont{Lehtinen}},
  \bibinfo{author}{\bibfnamefont{A.}~\bibnamefont{Paila}},
  \bibinfo{author}{\bibfnamefont{Y.}~\bibnamefont{Makhlin}}, \bibnamefont{and}
  \bibinfo{author}{\bibfnamefont{P.}~\bibnamefont{Hakonen}},
  \bibinfo{journal}{Phys. Rev. Lett.} \textbf{\bibinfo{volume}{96}},
  \bibinfo{pages}{187002} (\bibinfo{year}{2005}).

\bibitem[{\citenamefont{Wilson et~al.}(2007)\citenamefont{Wilson, Duty,
  Persson, Sandberg, Johansson, and Delsing}}]{Wilson:LZinterf}
\bibinfo{author}{\bibfnamefont{C.~M.} \bibnamefont{Wilson}},
  \bibinfo{author}{\bibfnamefont{T.}~\bibnamefont{Duty}},
  \bibinfo{author}{\bibfnamefont{F.}~\bibnamefont{Persson}},
  \bibinfo{author}{\bibfnamefont{M.}~\bibnamefont{Sandberg}},
  \bibinfo{author}{\bibfnamefont{G.}~\bibnamefont{Johansson}},
  \bibnamefont{and} \bibinfo{author}{\bibfnamefont{P.}~\bibnamefont{Delsing}},
  \bibinfo{journal}{Phys. Rev. Lett.} \textbf{\bibinfo{volume}{98}},
  \bibinfo{pages}{257003} (\bibinfo{year}{2007}).

\bibitem[{\citenamefont{Izmalkov et~al.}(2008)\citenamefont{Izmalkov, van~der
  Ploeg, Shevchenko, Grajcar, H{\"u}bner, Omelyanchouk, and
  Meyer}}]{Izmalkov:LZinterf}
\bibinfo{author}{\bibfnamefont{A.}~\bibnamefont{Izmalkov}},
  \bibinfo{author}{\bibfnamefont{S.~H.~W.} \bibnamefont{van~der Ploeg}},
  \bibinfo{author}{\bibfnamefont{S.~N.} \bibnamefont{Shevchenko}},
  \bibinfo{author}{\bibfnamefont{M.}~\bibnamefont{Grajcar}},
  \bibinfo{author}{\bibfnamefont{U.}~\bibnamefont{H{\"u}bner}},
  \bibinfo{author}{\bibfnamefont{A.~N.} \bibnamefont{Omelyanchouk}},
  \bibnamefont{and} \bibinfo{author}{\bibfnamefont{H.}~\bibnamefont{Meyer}},
  \bibinfo{journal}{Phys. Rev. Lett.} \textbf{\bibinfo{volume}{101}},
  \bibinfo{pages}{017003} (\bibinfo{year}{2008}).

\bibitem[{\citenamefont{Oliver et~al.}(2009)\citenamefont{Oliver, Yu, Lee,
  Berggren, Levitov, and Orlando}}]{Oliver:LZinterf}
\bibinfo{author}{\bibfnamefont{W.}~\bibnamefont{Oliver}},
  \bibinfo{author}{\bibfnamefont{Y.}~\bibnamefont{Yu}},
  \bibinfo{author}{\bibfnamefont{J.}~\bibnamefont{Lee}},
  \bibinfo{author}{\bibfnamefont{K.}~\bibnamefont{Berggren}},
  \bibinfo{author}{\bibfnamefont{L.}~\bibnamefont{Levitov}}, \bibnamefont{and}
  \bibinfo{author}{\bibfnamefont{T.~P.} \bibnamefont{Orlando}},
  \bibinfo{journal}{Science} \textbf{\bibinfo{volume}{310}},
  \bibinfo{pages}{1653} (\bibinfo{year}{2009}).

\bibitem[{\citenamefont{Kayanuma}(1997)}]{Kayanuma:LZ}
\bibinfo{author}{\bibfnamefont{Y.}~\bibnamefont{Kayanuma}},
  \bibinfo{journal}{Phys. Rev. A} \textbf{\bibinfo{volume}{55}},
  \bibinfo{pages}{R2495} (\bibinfo{year}{1997}).

\bibitem[{\citenamefont{H{\"a}nggi}(1998)}]{Hanggi:floquet}
\bibinfo{author}{\bibfnamefont{P.}~\bibnamefont{H{\"a}nggi}}, in
  \emph{\bibinfo{booktitle}{Quantum transport and dissipation}}, edited by
  \bibinfo{editor}{\bibfnamefont{T.}~\bibnamefont{Dittrich}},
  \bibinfo{editor}{\bibfnamefont{P.}~\bibnamefont{H{\"a}nggi}},
  \bibinfo{editor}{\bibfnamefont{G.-L.} \bibnamefont{Ingold}},
  \bibinfo{editor}{\bibfnamefont{B.}~\bibnamefont{Kramer}},
  \bibinfo{editor}{\bibfnamefont{G.}~\bibnamefont{Sch{\"o}n}},
  \bibnamefont{and} \bibinfo{editor}{\bibfnamefont{W.}~\bibnamefont{Zwerger}}
  (\bibinfo{publisher}{Wiley-VCH}, \bibinfo{address}{Weinheim},
  \bibinfo{year}{1998}).

\bibitem[{\citenamefont{Cheianov and Fal'ko}(2006)}]{CheianovFalko}
\bibinfo{author}{\bibfnamefont{V.~V.} \bibnamefont{Cheianov}} \bibnamefont{and}
  \bibinfo{author}{\bibfnamefont{V.~I.} \bibnamefont{Fal'ko}},
  \bibinfo{journal}{Phys. Rev. B} \textbf{\bibinfo{volume}{74}},
  \bibinfo{pages}{041403(R)} (\bibinfo{year}{2006}).

\bibitem[{\citenamefont{Shytov et~al.}(2008)\citenamefont{Shytov, Gu, and
  Levitov}}]{Shytov:LZmagn}
\bibinfo{author}{\bibfnamefont{A.}~\bibnamefont{Shytov}},
  \bibinfo{author}{\bibfnamefont{N.}~\bibnamefont{Gu}}, \bibnamefont{and}
  \bibinfo{author}{\bibfnamefont{L.~S.} \bibnamefont{Levitov}},
  \bibinfo{journal}{arXiv e-print}  (\bibinfo{year}{2008}), \eprint{0708.3081}.

\bibitem[{\citenamefont{Yampol'skii et~al.}(2008)\citenamefont{Yampol'skii,
  Savel'ev, and Nori}}]{YampNori:oscilcond1}
\bibinfo{author}{\bibfnamefont{V.~A.} \bibnamefont{Yampol'skii}},
  \bibinfo{author}{\bibfnamefont{S.}~\bibnamefont{Savel'ev}}, \bibnamefont{and}
  \bibinfo{author}{\bibfnamefont{F.}~\bibnamefont{Nori}}, \bibinfo{journal}{New
  J. Phys} \textbf{\bibinfo{volume}{10}}, \bibinfo{pages}{053024}
  (\bibinfo{year}{2008}).

\end{thebibliography}
\end{document}